\documentclass[twocolumn,showpacs,pre,floatfix]{revtex4}
\usepackage{amsmath,amsfonts,amssymb}
\usepackage{graphicx} 

\begin{document}
\title{Edwards-like statistical mechanical description of the parking
lot model for vibrated granular materials.}
\author{G. Tarjus and P. Viot} \
\affiliation{Laboratoire de Physique
Th{\'e}orique des Liquides, Universit{\'e} Pierre et Marie Curie, 4, place
Jussieu,\\ 75252 Paris Cedex, 05 France}

\begin{abstract}
We   apply the statistical  mechanical approach  based on the ``flat''
measure proposed by Edwards and coworkers  to the parking lot model, a
model  that reproduces  the  main features   of the  phenomenology  of
vibrated granular materials. We  first build the  flat measure for the
case of vanishingly  small  tapping strength and then   generalize the
approach   to finite   tapping    strengths  by  introducing   a   new
``thermodynamic''  parameter,  the   available  volume  for   particle
insertion,  in addition to the  particle density.  This description is
able  to take  into account  the   various memory effects observed  in
vibrated  granular  media.  Although  not  exact, the approach gives a
good description  of  the behavior  of  the parking-lot  model  in the
regime of slow compaction.
\end{abstract}
\pacs{05.70.Ln,45.70.Cc}
\maketitle
\section{Introduction}
Granular media are a-thermal, out-of-equilibrium systems that it would
be  useful  to describe  within  a statistical mechanical framework. A
given macro-state of such a system characterized by a fixed density of
grains (consider for simplicity   a packing of monodisperse  spherical
particles) is very likely to be associated with an exponentially large
number of micro-states   or particle configurations.  How  the packing
was  prepared (by pouring,   shaking, shearing, etc...) may influence
its   properties  and  change     the  way  the  associated   particle
configurations  are sampled when  repeating over the same experimental
protocol.   However, the simplest  hypothesis,  put forward by Edwards
and his coworkers\cite{EO89,ME89,E94,EG99},  is that  all micro-states
characterized by a  given average  density are  equiprobable. With this  ``flat
measure'', on can  build a statistical  mechanical framework  in which
entropy, i.e.,  the logarithm of  the  number of  micro-states, is the
relevant thermodynamic potential.  This approach has recently been the
focus of an intense research activity,  in connection with a series of
experiments  performed    on      weakly    vibrated      granular
materials\cite{KFLJN95,NKBJN98,BKNJN98,JTMJ00,PB02}    and  with     a
theoretical description of  out-of-equilibrium glassy systems based on
the                    concept         of                    effective
temperature\cite{CKP97,CR03,K00,MakseK02,LN01}.

In the past few years, the Edwards' hypothesis has been tested on many
models, virtually  all of them being  lattice models with some kind of
``tapping''
kinetics\cite{BKLS00,BKLS01,DL01,LD01,LD01b,DL02b,L02,S02,CFN01,FNC02b,FNC02,NFC02,dMGL02,dMGL03,PBS00,PrB02,PrB03,BPr02,BFS02}. In
the   absence  of    experimental  tests   of   this  approach   (see,
however,\cite{NKBJN98}),  such theoretical studies  are expected to
better circumscribe  the conditions  of   validity of the  statistical
mechanical description.  (Presumably, only ``approximate validity'' can be
expected   since, aside  from specific  mean-field models\cite{CKP97,K00},
such a simplified  description  of  out-of equilibrium situations   in
terms of a small number of ``thermodynamic'' parameters is unlikely to
be exact.)

In  this article, we   consider an Edwards-like statistical mechanical
description   approach  for  the   one-dimensional  model    of random
adsorption-desorption of hard  particles\cite{TST90,JTT94,KB94},  also
known as the     parking-lot-model\cite{NKBJN98}. This latter     is a
microscopic,   off-lattice model  that mimics  many   features  of the
compaction  of a  vibrated  column  of  grains.  Besides a qualitative
description  of  the    phenomenology  of   weakly tapped     granular
media\cite{NKBJN98,BKNJN98,KNT99,TTV99,TTV00,TTV01,TTV03},         the
interest of the model is that exact analytical  results can be derived
or, when  not possible, very accurate numerical  data can  be obtained
from computer  simulations. In the  next section, we briefly introduce
the parking-lot  model and   we  discuss  its connection   to vibrated
granular   materials.  In section   III,   we consider  the  limit  of
vanishingly  small (but non-zero) tapping  intensity; we construct for
this case an Edwards-like description based on a flat measure over all
``blocked'' states and we  compare  the resulting predictions to   the
exact behavior.  In the following  section, we generalize the study to
the case of a finite tapping intensity: we consider what appears to be
the simplest, yet   compatible with  known experimental  observations,
generalization of the  Edwards'   formalism. Finally, we  discuss  the
merits and limitations of the approach.

\section{The model and its connection to vibrated granular materials} 
The  parking-lot       model     is    a      one-dimensional   random
adsorption-desorption process  of hard rods  on a line.  Hards rods of
length $\sigma$ are deposited at random positions on a line at rate $k_{+}$
and are effectively inserted if they do not overlap with pre-deposited
rods;   otherwise they  are  rejected.    In  addition, all  deposited
particles can desorb, i.e., be ejected from  the line at random with a
rate $k_{-}$.   Time is measured in units  of $1/k_+$, length in units
of   $\sigma$, and   the model  depends   on one  control   parameter  $K =
k_{+}/k_{-}$.  When no desorption   is  present ($k_-=0$), the   model
reduces to  the purely irreversible one-dimensional  random sequential
adsorption (RSA) process\cite{R63,E93,TTVV00}, also  known as  the car
parking problem, and all the properties of the system as a function of
time are  known exactly\cite{R63,BonnierBV94,VTT93}.  In addition, for
$1/K$ non strictly equal     to zero, the competition    of mechanisms
between adsorption and desorption allows the  system to reach a steady
state   that is nothing  but   an equilibrium fluid   of  hard rods at
constant activity $1/K$: there too, all properties are known exactly.

The densification kinetics of the parking-lot model at constant $K$ is
described by 
\begin{equation}\label{eq:1}
\left.\frac{\partial\rho }{\partial t}\right|_K=\Phi(t)-\frac{\rho}{K}
\end{equation}
where  $\rho(t)$ the  density of hard rods on the line at time $t$
(recall that $\sigma\equiv 1$) 
and  $\Phi (t)$ is the  fraction of the line that is available at time
$t$ for inserting a
new particle, i.e., the probability associated with finding  an
interval free of particles (a ``gap'') of length at least $1$. The
quantities $\rho$ and $\Phi$ can be calculated from the $1-$gap distribution
function $G(h,t)$ which is the density of gaps of length $h$ at time
$t$ via a number of ``sum rules'':
\begin{align}
\rho(t)&=\int_0^\infty dh G(h,t),\label{eq:2}\\
1-\rho(t)&=\int_0^\infty dh hG(h,t),\label{eq:3}\\
\Phi(t)&=\int_0^\infty dh(h-1)G(h,t).\label{eq:4}
\end{align}
The    evolution with time of  the   $1-$gap distribution function can
itself  be  described by   a kinetic equation   that involves  $2-$gap
distribution functions,   and so on\cite{TTV00}.  Except  for  the two
above mentioned limits (RSA   when $k_-=0$, equilibrium  when $t\to+\infty$),
the  infinite   hierarchy  of  coupled   equations  cannot   be solved
analytically and  one must resort  to approximate treatments and computer
simulations, as described in previous articles\cite{TTV99,TTV00,TTV01,TTV03}.

First introduced in the context of  protein adsorption at liquid-solid
interfaces\cite{TST90,JTT94,KB94},    the random adsorption-desorption
model has recently been applied to  the description of weakly vibrated
granular
materials\cite{NKBJN98,BKNJN98,KNT99,TTV99,TTV00,TTV01,TTV03}.     The
connection between  the parking-lot model and these  latter is made by
regarding the particles on the  line as an average  layer of grains in
the vibrated column.  Time measures the number of taps whose effect is
to eject particles from the layer; ejection is followed by the arrival
at random of particles  in the layer,  which mimics the gravity-driven
relaxation step   in   the experiment.   Considering  that   the  main
influence of the intensity of the tapping is  to determine the average
number  of particles  ejected at   each  tap,  (this  number being  an
increasing  function of intensity), leads to  associate $1/K$ with the
tapping strength.   A two-dimensional version of  the model  with some
polydispersity of the particles  would clearly be more  realistic, but
one does  not expect this to change   the qualitative features  of the
model\cite{TTVV00}.  A more serious  caveat is the absence of explicit
account    of  the mechanical  stability  of    the particle packings:
stability is only implicitly described by the  fact that the particles
are blocked on the line between two successive desorption events.

Despite its  drastic  simplification of  the situation  encountered in
vibrated granular  materials, the  parking  lot model reproduces at  a
qualitative level most of  the  relevant phenomenology: (i) for  large
rate $K$ corresponding  to weak tapping intensity, compaction proceeds
very slowly  and can be  effectively described by an inverse logarithm
of time\cite{JTT94,KB94,TTV99,TTV00}; (ii)  stronger tapping  leads to
faster   initial    compaction  but   to   less   effective asymptotic
packing\cite{TTV00};  (iii)  the slow densification  kinetics leads to
irreversibility  effects and to the observation  of two curves for the
packing  density  as a function   of tapping intensity\cite{TTV00} one
essentially   reversible   and   one  irreversible  depending   on the
experimental protocol chosen\cite{TTV01};  (iv) the power spectrum  of
the density      fluctuations near the   steady  state   is distinctly
non-Lorenztian  and  displays   a  power-law   regime at  intermediate
frequencies\cite{NKBJN98,BKNJN98,KNT99,TTV01,TTV03};  (v)  non-trivial
memory effects   are   observed when changing    abruptly  the tapping
intensity\cite{TTV01,TTV03}.

Note that, as recently studied for a  one-dimensional model with tapping
dynamics\cite{PrB03}, the  parking  lot model  could
also  be  used,  via the introduction   of  two kinds of particles,  to
describe  the  segregation   phenomena   with  the  so-called   Brazil
Nut\cite{RSPS87,MLNJ01} and  Reverse Brazil Nut\cite{HQL01,BEKR03}
effects.

\section{Limit of vanishingly small tapping intensity: $K\to\infty$}
In this  limit, ejection of one particle  from the line is followed by
an infinite number  of insertion trials until  one, or seldom, two new
particles are added. The stable or ``blocked'' configurations are thus
those for which no more  particle insertions are possible (recall that
once successfully inserted particles cannot  move on the line),  i.e.,
all configurations of non-overlapping  rods  such that the   available
line fraction  $\Phi$  is  zero, or, equivalently,   such  that  all gaps
between neighboring particles are smaller than a particle size (here
taken     as unity).   Edwards'   prescription    for  constructing  a
statistical-mechanical description of  this system is then to consider
that all such  ``blocked'' configurations at a  fixed density $\rho$  are
equiprobable.

Consider a line of length $L$ with $N$ particles. With periodic
boundary conditions, this system has also $N$ gaps between neighboring
particles. Denoting by $h_1,h_2,\ldots,h_N$ the lengths of these gaps, the
total number of ``blocked'' configurations is given by the
configurational integral calculated under the constraint that $h_i<1$
for $i=1,..,N$, namely,
\begin{align}\label{eq:5}
Z(L,N)=&\int_0^1\ldots\int_0^1\left(\prod_{i=1}^Ndh_i\right)\delta(L-N-\sum_{i=1}^{N}h_i),
\end{align}
which by using the integral representation of the delta-function, can
be rewritten as
\begin{align}
Z(L,N)=&\int_Cdze^{z(L-N)}\left(\prod_{i=1}^N\int_0^1dh_i e^{-zh_i}\right),
\end{align}
where $C$ denotes a closed contour. Integrating over the $h_i'$s
yields
\begin{align}\label{eq:6}
Z(L,N)=\int_Cdz\exp\left(L\left[z(1-\rho)+\rho
\ln\left(\frac{1-\exp(-z)}{z}\right) \right]\right),
\end{align}
where $\rho=N/L$. In the macroscopic limit where $N\to\infty$, $L\to\infty$, with
fixed $\rho$, the above expression can be evaluated though a saddle-point
method, which gives
\begin{equation}
Z(L,N)\simeq \exp(Ls(\rho))
\end{equation}
where  $s(\rho)$, the entropy density, is expressed as
\begin{equation}
s(\rho)=(1-\rho)z+\rho \ln\left(\frac{1-e^{-z}}{z}\right)
\end{equation}
where $z\equiv z(\rho)$ is the solution of the saddle-point equation
\begin{equation}\label{eq:7}
\left(\frac{1-\rho }{\rho}\right)=\frac{1}{z}-\frac{e^{-z}}{1-e^{-z}}.
\end{equation}

In    Edwards'   language,     $\frac{1}{z}=\left(\left.\frac{\partial(Ls)}{\partial
L}\right|_N\right)^{-1}$  is  the    compactivity (up  to   a  trivial
constant)\cite{footn}. In an equilibrium system of hard rods, i.e., a flat measure
without the constraint  that all gaps have a  length smaller than $1$,
$z(\rho)$ would  simply be equal  to $P/(k_BT)=\rho/(1-\rho)$ where $P$  is the
pressure.

By Legendre transforming the entropy $S(L,N)$, one obtains a new
potential
\begin{align}
Y(N,z)=-S(L,N)+zL=N\left[z-\ln\left(\frac{1-e^{-z}}{z}\right)\right]
\end{align}
such that $\left.\frac{\partial(Y)}{\partial z}\right|_N=\langle L\rangle$ and from which
one can obtain the fluctuations of the system size,
\begin{align}
\langle L^2\rangle -\langle  L\rangle^2=-\left.\frac{\partial^2(Y)}{\partial z^2}\right|_N=-N\left.\frac{\partial(1/\rho)}{\partial z}\right|_N.
\end{align}
By  combining the above expression with Eq.~(\ref{eq:7}), one derives
the fluctuations of the density,
\begin{align}
L\left(\langle \rho^2\rangle -\langle \rho \rangle^2\right)=\frac{\rho^3}{N}\left(\langle
L^2\rangle -\langle  L\rangle^2\right)\nonumber\\
=\rho^3\left(\frac{1}{z^2}-\frac{e^{-z}}{(1-e^{-z})^2}\right).\label{eq:8}
\end{align}

In the Edwards-like ensemble, one can also calculate the gap
distribution functions. The $1$-gap distribution function $G(h;\rho)$,
which gives the density of gaps of length $h$, is obtained from
\begin{align}
&G_{Ed}(h;\rho)=\frac{1}{LZ(L,N)}\nonumber\\
&\sum_{i=1}^N\int_0^1...\int_0^1
\left(\prod_{j=1,j\neq   i}^Ndh_j\right)\delta \left(L-N-h-\sum_{j=1,j\neq i}^{N}h_j\right).
\end{align}
By using the same method as before one finds that
\begin{align}
G_{Ed}(h;\rho)=\begin{cases}
      \rho\displaystyle\frac{z}{1-e^{-z}}e^{-zh} & \mbox{for } h<1,\\
	0 & \mbox{for } h>1,
      \end{cases}
\end{align}
where $z$ is the solution of Eq.~(\ref{eq:7}). It is easy to check
that  the above expression satisfies the two sum rules,
Eqs.~(\ref{eq:2}) and (\ref{eq:3}).

The higher-order gap distribution functions are obtained along the
same lines, and they satisfy a factorization property analogous to
that found for an equilibrium system of hard rods, i.e.,
\begin{align}\label{eq:9}
G_{Ed}(h,h';\rho)=G_{Ed}(h;\rho)G_{Ed}(h';\rho)\\
G_{Ed}(h,h',h'';\rho)=G_{Ed}(h;\rho)G_{Ed}(h';\rho)G_{Ed}(h'';\rho),
\label{eq:10}
\end{align}
etc.

The $1$-gap distribution function $G(h;\rho)$ is directly related to the
nearest-neighbor pair distribution function that represents the probability of
finding two neighboring particles whose centers are separated by a
distance $1+h$. It is also possible to calculate the full pair
distribution function $g(r;\rho)$ via a method which closely follows that
developed for the equilibrium system of hard rods\cite{F64}. The steps of the
calculation are detailed in Appendix A, and the final result reads

\begin{align}\label{eq:11}
g_{Ed}(r;\rho)=&\frac{1}{\rho}\sum_{m=1}^{\infty } \frac{(r-m)^{m-1}}{(m-1)!}  \left(
\sum_{k=0}^mC_m^k(-1)^k\theta(r-m-k)\right)\nonumber\\
&\left( \frac{z}{1-e^{-z}} \right)^m e^{-z(r-m)} 
\end{align} 
 where $\theta  (x)$  is  the Heaviside step function  and $z$ is the solution
of Eq.~(\ref{eq:7}). For comparison, we give the equilibrium pair
distribution function\cite{F64},
\begin{align}
g_{eq}(r;\rho)=&\frac{1}{\rho}\sum_{m=1}^\infty \theta(r-m)
\frac{(r-m)^{m-1}}{(m-1)!}
\nonumber\\&\left(\frac{\rho}{1-\rho}\right)^m
e^{\left(-\left(\frac{
\rho }{1-\rho}\right)(r-m)\right)}
\end{align}
\begin{figure}
\resizebox{8cm}{!}{\includegraphics{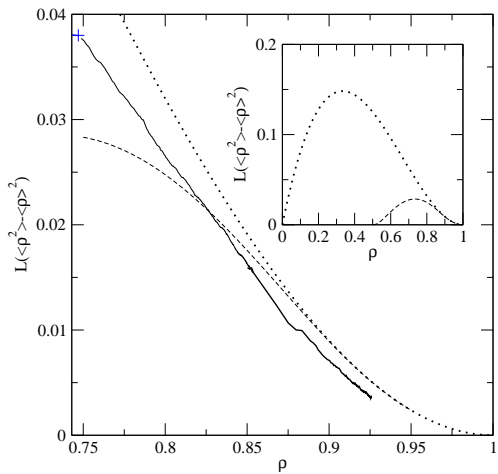}}
\caption{Density fluctuations of the parking lot model when $K\to+\infty $
for densities above the RSA  jamming limit.  Simulation data are shown
as the full line, the  Edwards approximation corresponds to the dashed
curve,  and  the equilibrium result  to the   dotted curve.  The inset
displays  the density fluctuations at equilibrium   and in the Edwards
approximation for a larger range of densities.}\label{fig:1}
\end{figure}

\begin{figure}[t]
\resizebox{8cm}{!}{\includegraphics{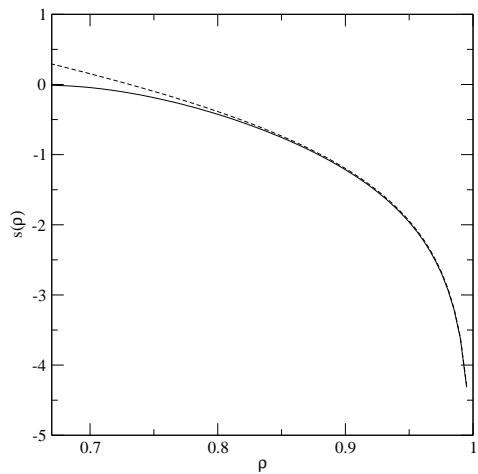}}
\caption{Entropy density versus particule density for a hard-rod system: Edwards
entropy   for  the parking lot  model   when  $K\to\infty$  (full  curve) and
equilibrium entropy (dashed curve).}\label{fig:2}
\end{figure}

We can now compare  the above results  derived under the  condition of
equiprobability of the  ``blocked'' configurations with the exact ones
obtained either analytically or  numerically. When $K\to\infty $,  analytical
results are available in the two limits, $t=0^+$, which corresponds to
the purely irreversible RSA process at the jamming limit where no more
particles can be inserted\cite{R63}, and $t\to+\infty$ which corresponds to a
close-packed state with $\rho=1$.

For  the RSA at the jamming limit, closed-form expressions have been
derived for the saturation density\cite{R63},
\begin{align}
\rho_{JL}=\int_0^\infty dt \exp\left(-2\int_0^tdu\frac{1-e^{-u}}{u}\right)\simeq 0.74759\ldots,
\end{align}
for   the   density  fluctuations\cite{BonnierBV94},   for  the    gap
distribution  functions\cite{VTT93}  and   for  the pair  distribution
function\cite{BonnierBV94}.  (The expressions  are given in Appendix B.)
When comparing  to  the  Edwards-like  results  at  the  same  density,
$\rho_{JL}$, one  finds qualitative differences.    Most notably, (i) the
exact $1$-gap distribution  function displays a logarithmic divergence
at contact between particles ($h\to0^+$),
\begin{equation}
G(h;\rho_{JL})\simeq- e^{-2\gamma}\ln(h)
\end{equation}
where $\gamma$ is the Euler constant, (ii) the exact multi-gap distribution
functions do not reduce to  products  of the $1$-gap functions,  (iii)
the exact pair distribution function  has a super-exponential decay at
large distances,
\begin{align}
g(r,\rho_{JL})-1\sim \frac{1}{\Gamma(r)}&,\,\,\,\,r\to\infty
\end{align}
all features that are missed by the flat-measure expressions, since
\begin{align}
G_{Ed}(h=0;\rho_{JL})=\rho_{JL}\frac{z(\rho_{JL})}{1-z(\rho_{JL})}\\
g_{Ed}(r,\rho_{JL})-1\sim -e^{z(\rho_{JL})r},\,\,\,\, r\to\infty 
\end{align}
and the multi-gap functions satisfy a factorization property,
Eqs.~(\ref{eq:9})-~(\ref{eq:10}).

Quantitatively, one can also see differences, e.g., in the density
fluctuations, the exact result at jamming being $L(\langle
 \rho^2\rangle-\langle \rho \rangle^2)\simeq 0.038 $ to be compared to  $L(\langle \rho^2\rangle-\langle \rho \rangle^2)_{Ed}\simeq 0.028 $.

\begin{figure}
\resizebox{8cm}{!}{\includegraphics{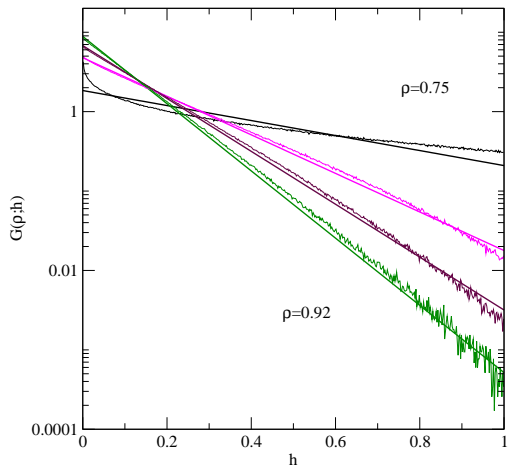}}
\caption{Parking lot model when $K\to+\infty$: log-linear plot of the $1$-gap
distribution function versus $h$ for several densities: from top to bottom
 $\rho=0.75,0.85, 0.89,0.91$.  Comparison of simulation data (wavy line) and
Edwards approximation (full curve).}\label{fig:3}
\end{figure}

\begin{figure}
\resizebox{8cm}{!}{\includegraphics{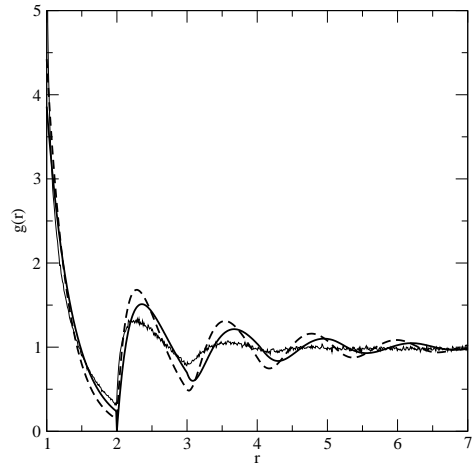}}

\caption{Pair distribution function in the parking lot model when
$K\to+\infty$ at  $\rho=0.77$:  simulation data  (wavy curve),   Edwards approximation
(full curve) and equilibrium (dashed curve).}\label{fig:4}
\end{figure}

\begin{figure}
\resizebox{8cm}{!}{\includegraphics{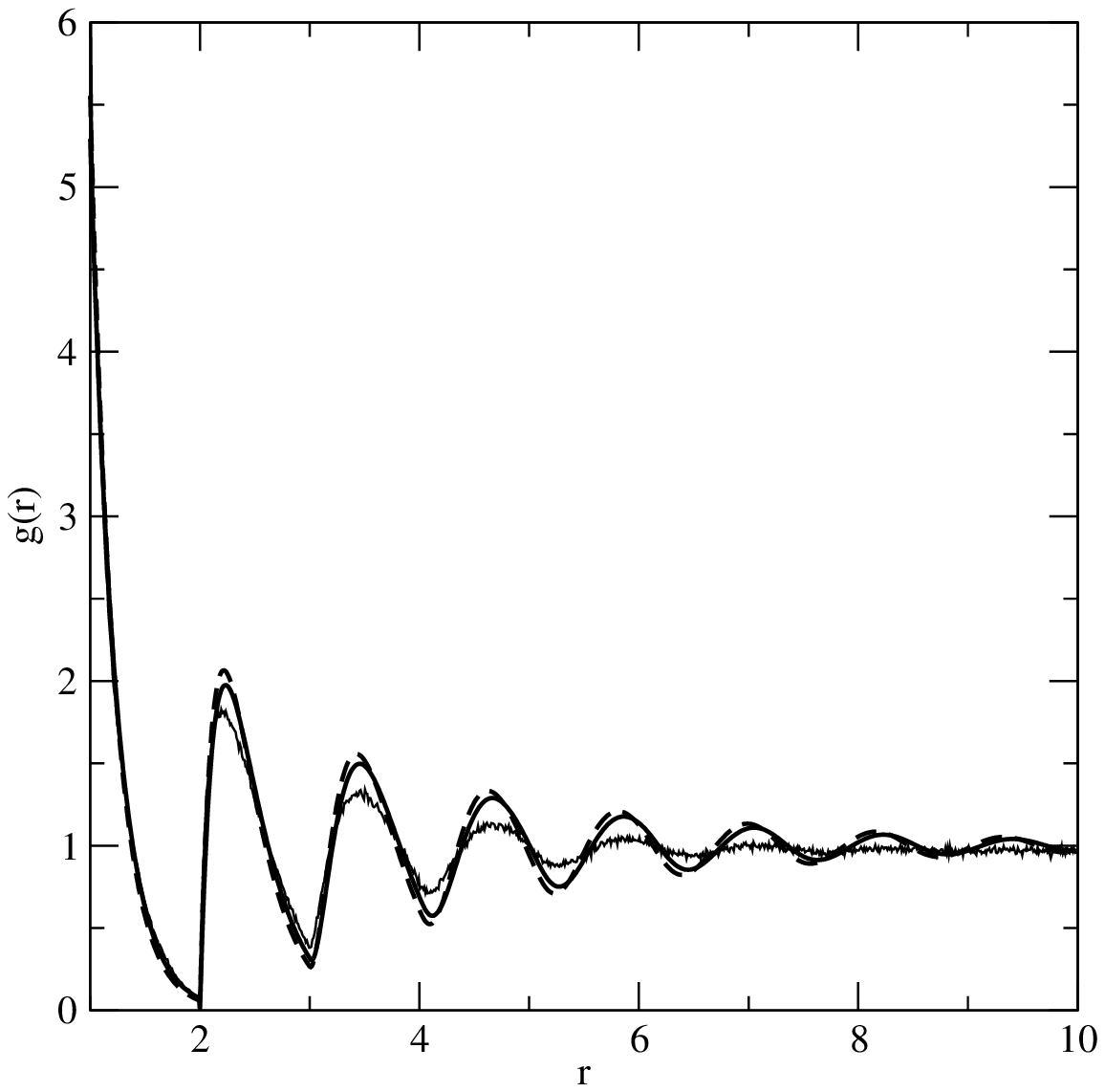}}
\caption{Same as Fig.~\ref{fig:4} for $\rho=0.82$}\label{fig:5}
\end{figure}

\begin{figure}
\resizebox{8cm}{!}{\includegraphics{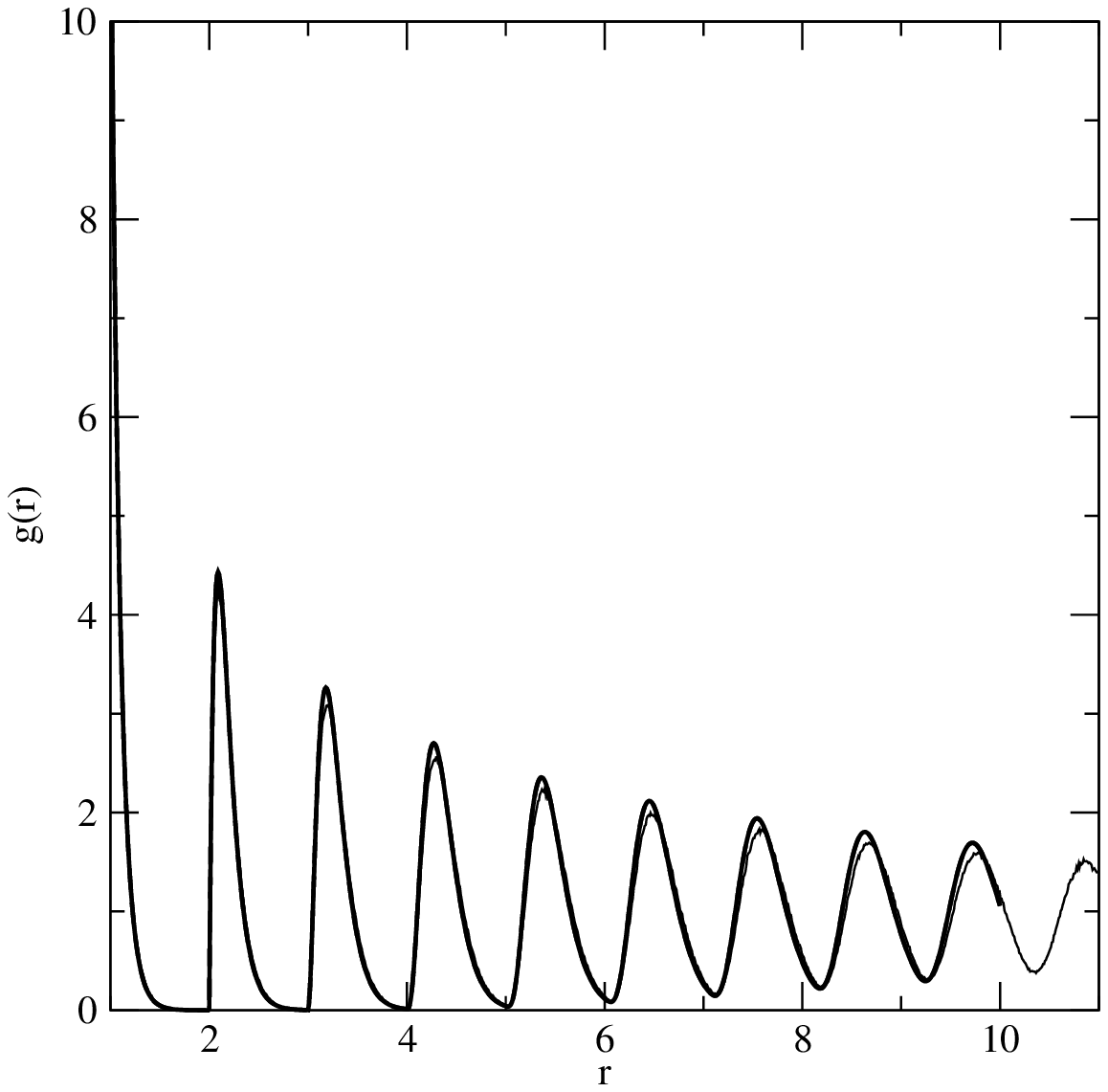}}
\caption{Same as Fig.~\ref{fig:4} for $\rho=0.92$}\label{fig:6}
\end{figure}

Such observations, that generalize to  an off-lattice model the results
obtained by de Smedt  {\it et al.}\cite{dMGL02,dMGL03} for  random and
cooperative sequential adsorption  models on a one-dimensional lattice,
are in fact to  be expected: it has   been shown that the  RSA process
generates configurations  of hard  particles  that are  sampled from a
probability distribution  that is ``biased''  when compared  to a flat
measure\cite{W66,TTS91}.

As the system further evolves  with time at vanishingly small  tapping
intensity,  compaction takes place  beyond the RSA saturation density,
and  the difference between  the actual properties   of the system and
those predicted by the Edwards-flat measure diminishes.

Figure~\ref{fig:1} compares the  variation with $\rho$ (for  $\rho\geq \rho_{JL}$)
of   the  density fluctuations of the    parking lot model (simulation
results) with   the  flat-measure result,   Eq.~(\ref{eq:8}), and  the
equilibrium   curve,  $L\left(\langle\rho^2\rangle-\langle \rho\rangle^2\right)_{eq}=\rho(1-\rho)^2$.  The
flat-measure prediction is in fair agreement with the simulation data,
especially for intermediate densities,  but it reaches too rapidly the
equilibrium curve     that   (slightly)   overestimates  the     exact
result.  (Note that simulating  the system becomes very time consuming
as  compaction goes on and, in practice, we cannot  go beyond a density
of about   $0.93$;   the infinite-time  limit   should   of  course be
$\rho=1$). However the Edwards  description improves upon the equilibrium
curve and gives the proper shape of the density dependence with
an inflection   around $\rho\simeq 0.87$.  The  closeness of  the flat measure
result and of the equilibrium one (the former being of course with the
constraint that no gaps have a length larger than $1$) at high density
is illustrated in Fig.~\ref{fig:2} where  we show the  entropy density
versus particle density.

In Figure~\ref{fig:3} we have  displayed on a logarithmic-linear  plot
the predicted and computed   $1$-gap distribution functions 
for four different densities. The Edwards approximation is too small
at small $h$ for $\rho=0.75$, which is reminiscent of the missing the
logarithm divergence at the RSA jamming limit
$\rho_{JL}=0.747$ (see above). Otherwise the agreement with the simulation data is
very good (recall that this is a log-linear plot). For $h>1$, the
Edwards approximation is exact since $G_{Ed}(h>1)=0$.

Finally, we have plotted in Figures~\ref{fig:4}-~\ref{fig:6} the pair
distribution functions (simulation data, Edwards approximation,
equilibrium curve) for three different densities, $\rho=0.77$, $\rho=0.82$
and $\rho=0.92$. The (constrained) flat measure is an improvement upon
equilibrium curve, but even at $\rho=0.82$ (Fig.~\ref{fig:5}), it
somewhat overestimates the oscillations at large distances. For
$\rho=0.92$ (Fig.~\ref{fig:6}), the difference between the three curves
is barely visible. Note that for $1\leq r<2$, the pair distribution
function is equal to the nearest-neighbor distribution function, hence
to the $1$-gap distribution function shown in Fig.~\ref{fig:3}

\section{Finite tapping intensity: $K$ finite}
\begin{figure}
\resizebox{8cm}{!}{\includegraphics{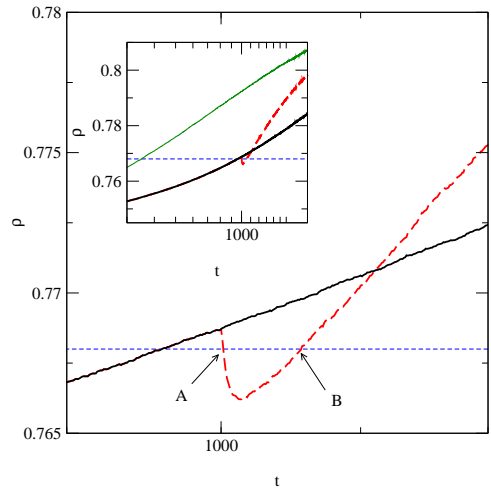}}
\caption{Memory effect in the parking lot model. The full curve
corresponds to a process at constant $K=2000$ whereas the dashed curve shows
the kinetics when $K$ is switched from $2000$ to $500$ at
$t_s=1000$. The points $A$ and $B$ corresponds to states with equal
density, equal value of $K=500$, but different further evolution. The
inset shows the phenomenon with a larger scale (upper curve: constant $K=500$,
lower curve: constant $K=2000$).}\label{fig:7}
\end{figure}

When the  tapping intensity, i.e., $1/K$,  is finite, one  must modify
the definition of the stable or ``blocked'' states. Indeed, for finite
$K$,      the  configuration  of    hard    rods    obtained   after a
desorption-adsorption event are no  longer characterized by $\Phi=0$, and
the gaps between particles can be larger than one. One also knows that
the  tapping  intensity $1/K$  is   not the  proper  ``thermodynamic''
parameter  to add   to  the density    in order  to   characterize the
macro-state of the system in   an Edwards-like statistical  mechanical
approach: the memory  effect  observed experimentally\cite{JTMJ00} and
reproduced by the  present parking lot model  (see above) implies that
the system can  be found in states characterized  by  the same density
$\rho$ and  the   same tapping  intensity   $1/K$   that however   evolve
differently  under further tapping  with the same intensity, $1/K$; as
illustrated in Fig.~\ref{fig:7}, the  density may increase in one case
and decrease in another.

If  the    tapping intensity  is   not  an   appropriate thermodynamic
parameter, a natural choice for a two-parameter statistical mechanical
description appears to  be $\Phi$, the available  line fraction, that one
can use   in conjunction   with the   density $\rho$.    A   non-zero $\Phi$
generalizes to a finite  tapping intensity, the prescription  uses for
vanishingly  small intensity  namely $\Phi=0$,  and $\Phi$  is also directly
relevant  for  describing  the   compaction   kinetics, as  shown   by
Eq.~(\ref{eq:1}).  We  thus consider a statistical mechanical ensemble
in which all configurations of non-overlapping hard rods characterized
by fixed values of $\rho$ and $\Phi$ are equally probable.

Denoting by $A$ the total length available for insertion for a
particle center ($A=\Phi L$), the configurational integral with the
constraints of fixed $A$, fixed system size $L$, and fixed number of
particles $N$ is given by

\begin{align}\label{eq:12}
Z(L,N,A)=&\int_0^1\ldots \int_0^1\prod_{i=1}^Ndh_i\delta\left(L-N-\sum_{i=1}^{N}h_i\right)\nonumber\\
&\delta\left(A-\sum_{i=1}^{N}\theta(h_i-1)(h_i-1)\right),
\end{align}
which can be rewritten as before as
\begin{align}
&Z(L,N,A)=\int_Cdz\int_{C'}dy\nonumber\\
&\exp\left(L\left[z(1-\rho)+y\Phi+\rho
\ln\left(\frac{z+y(1-\exp(-z))}{z(z+y)}\right) \right]\right)
\end{align}
where $C$ and $C'$ denote two  closed contours. In the macroscopic
limit, $N\to\infty$, $L\to\infty$, $A\to\infty$ with $\rho$ and $\Phi$ fixed, one can again use a
saddle point method to evaluate the integrals, which leads to 
\begin{equation}
Z(L,N,A)\simeq \exp(Ls(\rho,\Phi))
\end{equation}
where $s(\rho,\Phi)$ is expressed as
\begin{equation}
s(\rho,\Phi   )=(1-\rho)z+y\Phi +\rho  \ln\left(\frac{z+y(1-\exp(-z))}{z(z+y)}\right),
\end{equation}
with $z\equiv z(\rho,\Phi)$ and $y\equiv y(\rho,\Phi)$ solutions of the two coupled
equations
\begin{align}
\left(\frac{1-\rho }{\rho}\right)&=\frac{1}{z}+\frac{1}{z+y}-\frac{1+ye^{-z}}
{z+y(1-e^{-z})},\label{eq:13}\\
\frac{\Phi}{\rho}&=\frac{1}{z+y}-\frac{1-e^{-z}}
{z+y(1-e^{-z})}.\label{eq:14}
\end{align}
$z=\left(\frac{\partial S}{\partial L}\right)_{N,A}$ can again be considered as the inverse of
the ``compactivity'', but an additional intensive parameter,
$y=\left(\frac{\partial S}{\partial A}\right)_{N,L}$, is needed.

A double Legendre transformed potential $Y(N,z,y)$ can be introduced
as
\begin{align}
Y(N,z,y)&=-Ls(\rho,\Phi)+zL+yA\nonumber\\
&=N\left[z- \ln\left(\frac{z+y(1-\exp(-z))}{z(z+y)}\right)\right]
\end{align}
Then,  $\left.\frac{\partial(Y)}{\partial z}\right|_{N,y}=\langle L\rangle$ and
$\left.\frac{\partial(Y)}{\partial y}\right|_{N,z}=\langle A\rangle$, and the fluctuations in
$L$ and $A$ are given by
\begin{align}
\langle L^2\rangle -\langle  L\rangle^2&=-\left.\frac{\partial^2(Y)}{\partial
z^2}\right|_{N,y}=-N\left.\frac{\partial(1/\rho)}{\partial z}\right|_y\\
\langle A^2\rangle -\langle  A\rangle^2&=-\left.\frac{\partial^2(Y)}{\partial
y^2}\right|_{N,z}=-N\left.\frac{\partial(\Phi /\rho)}{\partial y}\right|_z\\
\langle LA\rangle -\langle L\rangle\langle A\rangle &=-\left.\frac{\partial^2(Y)}{\partial
z\partial y}\right|_{N}=-N\left.\frac{\partial(\Phi /\rho)}{\partial z}\right|_y\nonumber\\
&=-N\left.\frac{\partial(1/\rho)}{\partial y}\right|_z
\end{align}

By using the saddle-point equations, Eqs.~(\ref{eq:13}) and
(\ref{eq:14}), one arrives at the following expression for the
fluctuations of $\rho$,
\begin{align}
&L\left(\langle \rho^2\rangle -\langle \rho \rangle^2\right)=\frac{\rho^3}{N}\left(\langle
L^2\rangle -\langle  L\rangle^2\right)\nonumber\\
&=\rho^3\left(\frac{1}{z^2}+\frac{1}{(z+y)^2}-\frac{1+(2+z+y)ye^{-z}}{\left(z+y\left(1-e^{-z}\right)\right)^2}\right).
\end{align}

Similar expressions are obtained for the fluctuations of $\Phi$ and
the cross fluctuations of $\rho$ and $\Phi$, but are not shown here.

\begin{figure}[t]
\resizebox{8cm}{!}{\includegraphics{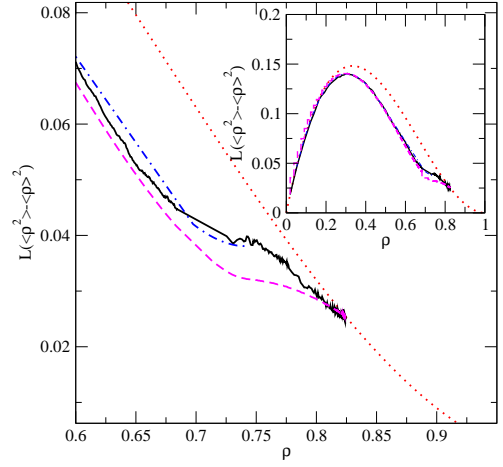}}
\caption{Density fluctuations as a function of $\rho$ for  $K=500$. The dotted
curve corresponds to the equilibrium density fluctuations, the dot-dashed
curve to the RSA result, the dashed curve to the $2$-parameter Edwards
prediction, and the wavy line to the simulation data. The inset displays
the same curves for a larger range of densities.}\label{fig:8}
\end{figure}
\begin{figure}[t]
\resizebox{8cm}{!}{\includegraphics{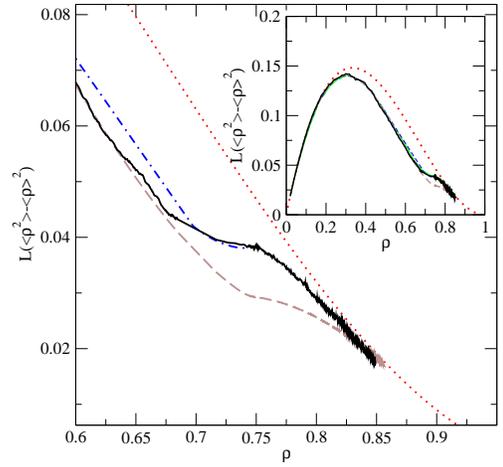}}
\caption{Same as Fig.~\ref{fig:8} for $K=5000$.}\label{fig:9}
\end{figure}

\begin{figure}
\resizebox{8cm}{!}{\includegraphics{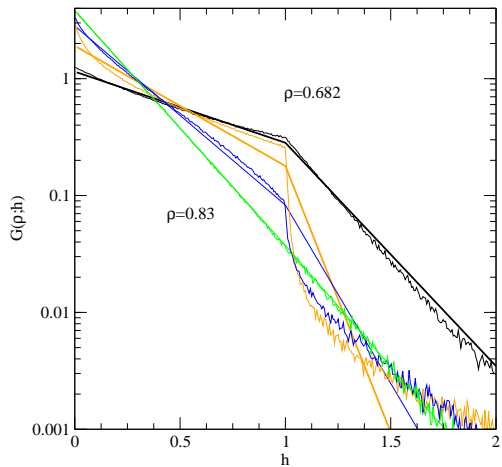}}
\caption{Parking lot model when $K=500$: log-linear plot of the $1$-gap
distribution function versus  $h$ for several  densities: from top  to
bottom in   the  middle  of the   figure  $\rho=0.682,0,748,0.790, 0.83$.
Comparison  of simulation data (wavy   line) and Edwards approximation
(full curve). For $\rho=0.83$, there is virtually no difference between
the two curves.}\label{fig:10}
\end{figure}
The gap distribution functions can also derived by following the same
method as in the previous section. This leads to
\begin{align}\label{eq:15}
G_{Ed}(h;\rho)=\begin{cases}
\rho\displaystyle\frac{z(z+y)}{z+y\left(1-e^{-z}\right)}e^{-zh}         &
\mbox{for                           }                           h<1,\\
\rho\displaystyle\frac{z(z+y)}{z+y\left(1-e^{-z}\right)}e^{-(zh+y(h-1))}
& \mbox{for } h>1,  \end{cases}
\end{align}
whereas the multi-gap distribution functions satisfy the factorization
property,       e.g.,   $G_{Ed}(h,h';\rho,\Phi )=G_{Ed}(h;\rho,\Phi )G_{Ed}(h';\rho,\Phi
)$. Notice that  the  $1$-gap  distribution  function is  a  piecewise
continuous   function     that   obeys     the   exact   sum    rules,
Eqs.~(\ref{eq:2})-~(\ref{eq:4}). It  is  also worth  pointing out that
the results of section  III ($K\to\infty $)  can  be recovered by taking  the
limit $y\to\infty$ in the   above equations. Finally, the  pair  distribution
function can be derived along  the same lines as  shown before and  in
Appendix  A, but the calculation is   too tedious and not sufficiently
insightful to be presented here.

A comparison between  the $2$-parameter Edwards  flat measure  and the
simulation data is shown in  Figs.~\ref{fig:8} and \ref{fig:9} for the
density fluctuations with $K=500$  and $K=5000$, respectively. We have
also   plotted the   equilibrium curve,  $L(\langle \rho^2\rangle_{eq}-\langle \rho\rangle_{eq}^2)=\rho
(1-\rho)^2$,  and the $1$-dimensional  RSA  curve\cite{BonnierBV94} up to
$\rho_{JL}$. The $2$-parameter flat measure predictions  are good but not
perfect.    They   display  the  proper    non-trivial   shape of  the
$\rho$-dependence,  in  contradistinction to the  equilibrium  curve, but
there  is a  significant underestimation of  the  fluctuations in the
density range around the RSA jamming limit.  Note that at high density
the flat measure  is in very good agreement  with  the simulation data
(at least   for  $K=500$) and does  not   merge too  rapidly  with the
equilibrium curve as seen above for the case $K\to\infty$.

The $1$-gap  distribution  functions  shown  on a log-linear   plot in
Fig.~\ref{fig:10} illustrate also the  overall good  agreement between
the  $2$-parameter    flat-measure predictions   and   the  simulation
data. The Edwards  approximation captures the change  of the slope  of
the $1$-gap   distribution function that  occurs   for $h>1$;  but the
curvature  seen  in the  simulation  data for $h$ slightly
larger than   $1$  is not correctly  reproduced   by the $2$-parameter
flat-measure which  predicts  an  exponential decay with a  factor
equal to $z+y$ (see Eq.~(\ref{eq:15})). Again, this discrepancy is
larger for densities around of the RSA jamming limit (the two
intermediate sets of curves in Fig.~\ref{fig:10}).

\section{Conclusion}
In this work we have applied the statistical mechanical approach based
on  the ``flat'' measure  proposed   by Edwards  and coworkers to  the
out-of-equilibrium situation obtained  in the parking-lot model.  This
latter is a microscopic  off-lattice  model that reproduces  the major
features of the phenomenology of vibrated  granular materials.  In the
statistical mechanical  description, a  macro-state  of the  system is
characterized by   fixed  values of two    macroscopic quantities, the
particle density $\rho$ and  the available line  fraction $\Phi$  (or rather
fixed values of $3$ extensive parameters, the number of particles, the
system  size  and   the  total   length available  for   insertion  of
particles), and all configurations  of non-overlapping  particles with
fixed $\rho$ and $\Phi$ are taken as equiprobable.

We   show   that such  an approach  misses    some  of the qualitative
signatures of the  limiting  case of a purely  irreversible adsorption
process  (RSA) at  the  jamming limit. However,   at higher densities,
i.e., in the slow (logarithmic)   compaction regime, it gives a  good,
yet    not  perfect,  quantitative   description   of many  observable
quantities (fluctuation, distribution functions).   The choice of $\Phi $
as an additional ``thermodynamic'' parameter  is  able to account  for
situations, encountered in various memory effects, in which macrostates
characterized by  the same density and the  same tapping intensity can
nonetheless be different.

The fact that a ``thermodynamic'' approach gives a good description of
a  model  of  vibrated   granular   media   is  promising.   In    the
one-dimensional  model  studied   here,  the generalized ``equation    of  state''
associated with the  flat measure can  be analytically derived so that
one  can  make theoretical predictions   concerning, e.g., the density
fluctuations  or the structure   of the  configurations.  However, one
still faces the task of predicting the state of the system for a given
preparation, i.e.,  for a given  time (number  of   taps) and a  given
protocol for the  tapping intensity. We  are presently working on this
problem.

\appendix

\section{Pair distribution function in the Edwards ensemble ($K\to\infty$)
and at equilibrium}
We introduce the probability density $\Psi_m(\xi)$ of finding two given
particles at a relative distance $\xi$, such that there is exactly $m-1$
particles between them; this  function can  be  expressed in  terms of
the 
partition function as
\begin{equation}
\Psi_m(\xi)=\frac{Z(\xi,m-1)Z(L-\xi,N-m)}{Z(L,N)}
\end{equation}
where $Z(L,N)$ can be calculated either with the Edwards measure or
at equilibrium.

For large $N$, one can use  the asymptotic expression of the partition
function   $Z(L,N)=e^{z(L-N)}\phi(z)^{N}$.   At    equilibrium $\phi(z)=1/z$
whereas  with the Edwards  measure, one gets $\phi(z)=(1-e^{(-z)})/z$ and
$z$  given  by Eq.~(\ref{eq:7}). Therefore,  the  probability  density
is equal to 
\begin{equation}
\Psi_m(\xi)=Z(\xi,m-1)\frac{1}{\phi(z)^{m}}e^{(-z(\xi-m))}
\end{equation}
and, formally, the pair distribution function can be expressed as
\begin{align}\label{eq:16}
g(r,\rho)=&\frac{1}{\rho}\sum_{m=1}^{\infty}\Psi_m (r)
\end{align}
For hard rods, the partition function $Z(\xi,m-1)$ is different from $0$
for $\xi >m$. At equilibrium, one has for $\xi>m$
\begin{equation}
Z(\xi,m-1)=\frac{(\xi-m)^{m-1}}{(m-1)!}
\end{equation} and $z=\rho/(1-\rho)$, which gives
for the pair distribution function\cite{F64} 
\begin{align}
g_{eq}(r,\rho)=&\frac{1}{\rho}\sum_{m=1}^\infty 
 \frac{\theta(r-m)(r-m)^{m-1}}{(m-1)!(1/\rho-1)^m}\exp\left(-\frac{r-m}{1/\rho-1}\right) 
\end{align}
 when $ r>1,$  where $\theta(x)$ is the Heaviside function.

In the Edwards ensemble, the expression for  $Z(\xi,m-1)$, hence for $\psi_m
(r )$ involves $m+1$ contributions. For instance, $\Psi_1  (r )$ is given
by
\begin{equation}
\Psi_1 (r )=(\theta(r-1)-\theta(r-2))\frac{z}{1-e^{-z}}e^{-z(r-1)}.
\end{equation}
More generally 
\begin{equation}\label{eq:17}
\psi_m (r )= \sum_{k=0}^mC_m^k(-1)^k\theta(r-m-k)\left(\frac{z}{1-e^{-z}}\right)^me^{-z(r-m)}
\end{equation}

Inserting    Eq.~(\ref{eq:17})   in    Eq.~(\ref{eq:16})     yields
Eq.~(\ref{eq:11})

\section{RSA expressions at the jamming limit}

For the one-dimensional RSA process, the $1$-gap distribution function
at the jamming limit ($t\to\infty$) is given by\cite{E93,TTVV00}
\begin{align}
G(h;\rho_{JL})=\begin{cases}
2\int_0^\infty dt t k(t)^2e^{(-th)}         &
\mbox{for }      h<1,\\
0& \mbox{for } h>1,  \end{cases}
\end{align}
with 
\begin{equation}
k(t)=\exp\left(-\int_0^t\frac{1-e^{-u}}{u}du\right).
\end{equation}
The  pair distribution function can be  expressed
in a closed form by using the Laplace transform $\tilde{g}(s,t)=
\displaystyle\int_0^{+ \infty} dl\,e^{-sl}g(l+1,t)$. At the jamming limit, one
has\cite{BonnierBV94} 

\begin{align}
\tilde{g}(s,\rho_{JL})=&\frac{1}{\rho^2_{JL}}\left(\frac{1}{s}\left[\int_0^\infty  dt_1\
\frac{k(t_1)k(t_1+s)}{k(s)}\right]^2\right.\nonumber\\
& -2\int_0^{\infty } dt_1\
\frac{k(t_1)k(t_1+s)}{k(s)} \int_0^{t_1}dt_2\
\frac{k(t_2)k(t_2+s)}{k(s)}\nonumber\\
&\left. \int_0^{t_2} dt_3\frac{e^{-t_3}k^2(s)}{k^2(t_3+s)} B(s,t_3)\right)\ .
\end{align}
with
\begin{equation}
B(s,t)=\frac{1}{s+t}-\frac{1-e^{-(s+t)}}{(s+t)^2}\ .
\end{equation}
The  expression for the density  fluctuations  follows from the  above
equation by taking the limit $s\to0$ of the expression $\rho\left(1+2\rho
\left(\tilde{g}(s,\rho_{JL})e^{-s}-\frac{1}{s}\right)\right)$\cite{BonnierBV94}. This gives
\begin{align}
L&\left(\langle \rho^2\rangle -\langle \rho \rangle^2\right)&=\rho_{JL}\left(-1 + 2\rho_{JL} 
- \frac{4}{\rho}\int_0^\infty  dt_1 k^2(t_1)\right.\nonumber\\& 
\int_0^{t_1} dt_2 k^2(t_2)&
\int_0^{t_2} dt_3 e^{-t_3} k^{-2}(t_3)\left.
\left(\frac{1}{t_3}-\frac{1-e^{-t_3}}{t_3^2}\right)\right),\nonumber\\
&\simeq 0.038.
\end{align}
At the same density, $\rho_{JL}$, the equilibrium value is $L\left(\langle \rho^2\rangle -\langle \rho \rangle^2\right)_{eq}=0.0476$.


\begin{thebibliography}{10}

\bibitem{EO89}
S. Edwards and R. Oakeshott, Physica A {\bf 157},  1080  (1989).

\bibitem{ME89}
A. Mehta and S. Edwards, Physica A {\bf 157},  1091  (1989).

\bibitem{E94}
S. Edwards,  in {\em Granular Matter: An Interdisciplinary Approach}, edited by
  A. Mehta (Springer-Verlag, New York, 1994).

\bibitem{EG99}
S. Edwards and D. Grinev, Phys. Rev. E {\bf 58},  4758  (1999).

\bibitem{KFLJN95}
J.~B. Knight, C.~G. Fandrich, C.~N. Lau, H.~M. Jaeger, and S.~R. Nagel, Phys.
  Rev. E {\bf 51},  3957  (1995).

\bibitem{NKBJN98}
E.~R. Nowak, J.~B. Knight, E. Ben-Naim, H.~M. Jaeger, and S.~R. Nagel, Phys.
  Rev. E {\bf 57},  1971  (1998).

\bibitem{BKNJN98}
E. Ben-Naim, J.~B. Knight, E.~R. Nowak, H.~M. Jaeger, and S.~R. Nagel, Physica
  D {\bf 123},  380  (1998).

\bibitem{JTMJ00}
C. Josserand, A. Tkachenko, D. Mueth, and H. Jaeger, Phys. Rev. Lett. {\bf 85},
   3632  (2000).

\bibitem{PB02}
P. Philippe and D. Bideau, Europhys. Lett. {\bf 60},  677  (2002).

\bibitem{CKP97}
L. Cugliandolo, J. Kurchan, and L. Peliti, Phys. Rev. E {\bf 55},  3898
  (1997).

\bibitem{CR03}
A. Crisanti and F. Ritort, J. Phys. A.: Math. Gen. {\bf 36},  R181  (2003).

\bibitem{MakseK02}
H.~A. Makse and J. Kurchan, Nature {\bf 415},  614  (2002).

\bibitem{LN01}
A. Liu and S. Nagel, {\em Jamming and Rheology: constrained dynamics} (Taylor
  and Francis, London, 2001).

\bibitem{K00}
J. Kurchan, J. Phys.: Condens. Matter {\bf 12},  6611  (2000).

\bibitem{BKLS00}
A. Barrat, J. Kurchan, V. Loreto, and M. Sellito, Phys. Rev. Lett. {\bf 85},
  503  (2000).

\bibitem{BKLS01}
A. Barrat, J. Kurchan, V. Loreto, and M. Sellito, Phys. Rev. E {\bf 63},
  051301  (2001).

\bibitem{DL01}
D. Dean and A. Lef{\`e}vre, Phys. Rev. Lett. {\bf 86},  5639  (2001).

\bibitem{LD01}
A. Lef{\`e}vre and D. Dean, Phys. Rev. E {\bf 64},  046110  (2001).

\bibitem{LD01b}
A. Lef{\`e}vre and D. Dean, J. Phys. A.: Math. Gen. {\bf 34},  L213  (2001).

\bibitem{DL02b}
D. Dean and A. Lef{\`e}vre, cond-mat/0212297  (2002).

\bibitem{L02}
A. Lef{\`e}vre, J. Phys. A.: Math. Gen. {\bf 35},  9037  (2002).

\bibitem{S02}
M. Sellitto, Phys. Rev. E {\bf 66},  042101  (2002).

\bibitem{CFN01}
A. Coniglio, A. Fierro, and M. Nicodemi, Physica A {\bf 302},  193  (2001).

\bibitem{FNC02b}
A. Fierro, M. Nicodemi, and A. Coniglio, Europhys. Lett {\bf 59},  642  (2002).

\bibitem{FNC02}
A. Fierro, M. Nicodemi, and A. Coniglio, Phys. Rev. E {\bf 66},  061301
  (2002).

\bibitem{NFC02}
M. Nicodemi, A. Fierro, and A. Coniglio, Europhys. Lett {\bf 60},  684  (2002).

\bibitem{PBS00}
A. Prados, J. Brey, and B. S{\'a}nchez-Rey, Physica A {\bf 284},  277  (2000).

\bibitem{PrB02}
A. Prados and J. Brey, Phys. Rev. E {\bf 66},  041308  (2002).

\bibitem{BPr02}
J. Brey and A. Prados, J. Phys: Condens. Matter {\bf 14},  1489  (2002).

\bibitem{PrB03}
A. Prados and J. Brey, cond-mat/03044654  (2003).

\bibitem{dMGL02}
G.~D. Smedt, C. Godr{\`e}che, and J. Luck, Eur. Phys. J. B {\bf 27},  363
  (2002).

\bibitem{dMGL03}
G.~D. Smedt, C. Godr{\`e}che, and J. Luck, Eur. Phys. J. B {\bf 32},  215
  (2003).

\bibitem{BFS02}
J. Berg, S. Franz, and M. Sellitto, Eur. Phys. J. B {\bf 26},  349  (2002).

\bibitem{JTT94}
X. Jin, G. Tarjus, and J. Talbot, J. Phys. A.: Math. Gen. {\bf 27},  L195
  (1994).

\bibitem{KB94}
P.~L. Krapivsky and E. Ben-Naim, J. Chem. Phys. {\bf 100},  6778  (1994).

\bibitem{TST90}
G. Tarjus, P. Schaaf, and J. Talbot, J. Chem. Phys. {\bf 93},  8352  (1990).

\bibitem{TTV00}
J. Talbot, G. Tarjus, and P. Viot, Phys. Rev. E {\bf 61},  5429  (2000).

\bibitem{TTV99}
J. Talbot, G. Tarjus, and P. Viot, J. Phys. A.: Math. Gen. {\bf 32},  2997
  (1999).

\bibitem{TTV01}
J. Talbot, G. Tarjus, and P. Viot, Eur. Phys. J. E {\bf 5},  445  (2001).

\bibitem{TTV03}
J. Talbot, G. Tarjus, and P. Viot, Fractals {\bf 11},  185  (2003).

\bibitem{KNT99}
A. Kolan, E. Nowak, and A. Tkachenko, Phys. Rev. E {\bf 59},  3094  (1999).

\bibitem{E93}
J. Evans, Rev. Mod. Phys. {\bf 65},  1281  (1993).

\bibitem{TTVV00}
J. Talbot, G. Tarjus, P.~V. Tassel, and P. Viot, Colloids and Surf. A: {\bf
  165},  287  (2000).

\bibitem{R63}
R{\'e}nyi, Sel. Trans. Math. Stat. Prob. {\bf 4},  205  (1963).

\bibitem{BonnierBV94}
B. Bonnier, D. Boyer, and P. Viot, J. Phys. A.: Math. Gen. {\bf 27},  3671
  (1994).

\bibitem{VTT93}
P. Viot, G. Tarjus, and J. Talbot, Phys. Rev. E {\bf 48},  480  (1993).

\bibitem{RSPS87}
A. Rosato, K. Strandburg, F. Prinz, and R. Swendsen, Phys. Rev. Lett. {\bf 58},
   1038  (1987).

\bibitem{MLNJ01}
M. M{\"o}bius, B. Lauerdale, S. Nagel, and H. Jaeger, Nature {\bf 414},  270
  (2001).

\bibitem{HQL01}
D. Hong, P. Quinn, and S. Luding, Phys. Rev. Lett. {\bf 86},  3423  (2001).

\bibitem{BEKR03}
A.~P.~J. Breu, H.-M. Ensner, C.~A. Kruelle, and I. Rehberg, Phys. Rev. Lett.
  {\bf 90},  014302  (2002).

\bibitem{footn}
If one interprets the line as a layer of grains in a vibrated colummn, $(1/z)$
  is not quite the "bulk" compactivity that would describe the whole system.
  Indeed the relevant $3-$D volume {\`a} la Edwards (for fixed total number of
  particles in the whole column) corresponds here to $L^2/N$ for $L$ fixed. The
  bulk compactivity would then be $X=\frac{1}{L^2}\left(\frac{\partial S
  }{\partial (1/N)}\right)_L$, which gives
  $X^{-1}=\rho^2\left(z+\ln\left(\frac{z}{1-e^{-z}}\right)\right)$.

\bibitem{F64}
I.~Z. Fisher, {\em Statistical Theory of Liquids} (The University of Chicago,
  Chicago and London, 1964).

\bibitem{W66}
B. Widom, J. Chem. Phys. {\bf 44},  3888  (1966).

\bibitem{TTS91}
G. Tarjus, P. Schaaf, and J. Talbot, J. Stat. Phys. {\bf 63},  167  (1991).

\end{thebibliography}

\end{document}